\documentclass[pdftex,citeautoscript,aps,pra,superscriptaddress,cite,amsmath,amsfonts,amssymb,twocolumn]{revtex4}
\renewcommand{\vec}[1]{\mathbf{#1}} 
\usepackage[pdftex]{graphicx}
\usepackage{color}
\usepackage{url}
\renewcommand{\Re}{\operatorname{Re}}
\renewcommand{\Im}{\operatorname{Im}}

\newcommand{\figref}[1]{Fig.~\ref{fig:#1}}

\newcommand{\eqnumref}[1]{(\ref{eq:#1})}
\renewcommand{\eqref}[1]{Eq.~\eqnumref{#1}}
\newcommand{\Eqref}[1]{Equation~\eqnumref{#1}}
\newcommand{\citeasnoun}[1]{Ref.~\onlinecite{#1}}
\newcommand{\secref}[1]{Sec.~\ref{sec:#1}}

\begin{document}
\title{Casimir forces in the time domain: Applications}

\author{Alexander~P. McCauley}
\affiliation{Department of Physics,
Massachusetts Institute of Technology, Cambridge, MA 02139}
\author{Alejandro~W. Rodriguez}
\affiliation{Department of Physics,
Massachusetts Institute of Technology, Cambridge, MA 02139}
\author{John D. Joannopoulos}
\affiliation{Department of Physics,
Massachusetts Institute of Technology, Cambridge, MA 02139}
\author{Steven G. Johnson}
\affiliation{Department of Mathematics,
Massachusetts Institute of Technology, Cambridge, MA 02139}

\begin{abstract}
Our preceding paper, ~\citeasnoun{RodriguezMc09:PRA}, introduced a
method to compute Casimir forces in arbitrary geometries and for
arbitrary materials that was based on a finite-difference time-domain
(FDTD) scheme. In this manuscript, we focus on the efficient
implementation of our method for geometries of practical interest and
extend our previous proof-of-concept algorithm in one dimension to
problems in two and three dimensions, introducing a number of new
optimizations.  We consider Casimir piston-like problems with
nonmonotonic and monotonic force dependence on sidewall separation,
both for previously solved geometries to validate our method and also
for new geometries involving magnetic sidewalls and/or cylindrical
pistons. We include realistic dielectric materials to calculate the
force between suspended silicon waveguides or on a suspended membrane
with periodic grooves, also demonstrating the application of PML
absorbing boundaries and/or periodic boundaries.  In addition we apply
this method to a realizable three-dimensional system in which a silica
sphere is stably suspended in a fluid above an indented metallic
substrate.  More generally, the method allows off-the-shelf FDTD
software, already supporting a wide variety of materials (including
dielectric, magnetic, and even anisotropic materials) and boundary
conditions, to be exploited for the Casimir problem.
\end{abstract}

\maketitle

\section{Introduction}
\label{sec:intro}

The Casimir force, arising due to quantum fluctuations of the
electromagnetic field~\cite{casimir}, has been widely studied over the
past few decades~\cite{Boyer74, Lamoreaux97, hochan1, Munday09} and
verified by many experiments. Until recently, most works on the
subject had been restricted to simple geometries, such as parallel
plates or similar approximations thereof. However, new theoretical
methods capable of computing the force in arbitrary geometries have
already begun to explore the strong geometry dependence of the force
and have demonstrated a number of interesting effects~\cite{Antezza06,
  Rodriguez07:PRL, Rodriguez08:PRL, Emig07, Emig07:ratchet,
  Rodriguez07:PRA, ReidRo09,Pasquali08,Pasquali09,gies06:edge}. A
substantial motivation for the study of this effect is due to recent
progress in the field of nano-technology, especially in the
fabrication of micro-electro-mechanical systems (MEMS), where Casimir
forces have been observed~\cite{Serry98} and may play a significant
role in ``stiction'' and other phenomena involving small surface
separations. Currently, most work on Casimir forces is carried out by
specialists in the field. In order to help open this field to other
scientists and engineers, such as the MEMS community, we believe it
fruitful to frame the calculation of the force in a fashion that may
be more accessible to broader audiences.

In~\citeasnoun{RodriguezMc09:PRA}, with that goal in mind, we
introduced a theoretical framework for computing Casimir forces via
the standard finite-difference time-domain (FDTD) method of classical
computational electromagnetism~\cite{Taflove00} (for which software is
already widely available). The purpose of this manuscript is to
describe how these computations may be implemented in higher
dimensions and to demonstrate the flexibility and strengths of this
approach. In particular, we demonstrate calculations of Casimir forces
in two- and three-dimensional geometries, including three-dimensional
geometries without any rotational or translational
symmetry. Furthermore, we describe a harmonic expansion technique that
substantially increases the speed of the computation for many systems,
allowing Casimir forces to be efficiently computed even on single
computers, although parallel FDTD software is also common and greatly
expands the range of accessible problems.

Our manuscript is organized as follows: First, in \secref{mult-exp},
we briefly describe the algorithm presented
in~\citeasnoun{RodriguezMc09:PRA} to compute Casimir forces in the
time domain. This is followed by an important modification involving a
harmonic expansion technique that greatly reduces the computational
cost of the method. Second, \secref{num-imp} presents a number of
calculations in two- and three-dimensional geometries. In particular,
\secref{2d-geoms} presents calculations of the force in the
piston-like structure of~\citeasnoun{Rodriguez07:PRL}, and these are
checked against previous results. These calculations demonstrate both
the validity of our approach and the desirable properties of the
harmonic expansion. In subsequent sections, we demonstrate
computations exploiting various symmetries in three dimensions:
translation-invariance, cylindrical symmetry, and periodic boundaries.
These symmetries transform the calculation into the solution of a set
of two-dimensional problems. Finally in~\secref{full-3d} we
demonstrate a fully three-dimensional computation involving the stable
levitation of a sphere in a high-dielectric fluid above an indented
metal surface.  We exploit a freely available FDTD
code~\cite{Farjadpour06}, which handles symmetries and cylindrical
coordinates and also is scriptable/programmable in order to
automatically run the sequence of FDTD simulations required to
determine the Casimir force~\cite{casimir-wiki}.  Finally, in the
Appendix, we present details of the derivations of the harmonic
expansion and an optimization of the computation of $g(t)$.

%\section{Brief intro to FDTD}
%local Hamiltonian - very general.  We can do subpixel smoothing for
%non-dispersive [CITE - ardavan 2006], anisotropic [CITE - Ardavan
%  2009], and dispersive [CITE - ???] to get 2nd order accuracy in grid
%discritization.  

%This serves as a demonstration of the flexibility of our FDTD
%code~\endnote{ab-initio.mit.edu/wiki/index.php/Casimir calculations in Meep},
%which can readily handle a variety of different boundary conditions,
%including cylindrical boundary conditions, without any
%modifications. We present a number of nontrivial computations with
%perfect metal geometries. Finally, we leave some of the details of the
%derivations of the harmonic expansion, etcetera, to the Appendix.

\section{Harmonic Expansion}
\label{sec:mult-exp}

In this section we briefly summarize the method
of~\citeasnoun{RodriguezMc09:PRA}, and introduce an additional step
which greatly reduces the computational cost of running simulations in
higher dimensions.

\begin{figure}[tb]
\includegraphics[width=0.3\textwidth]{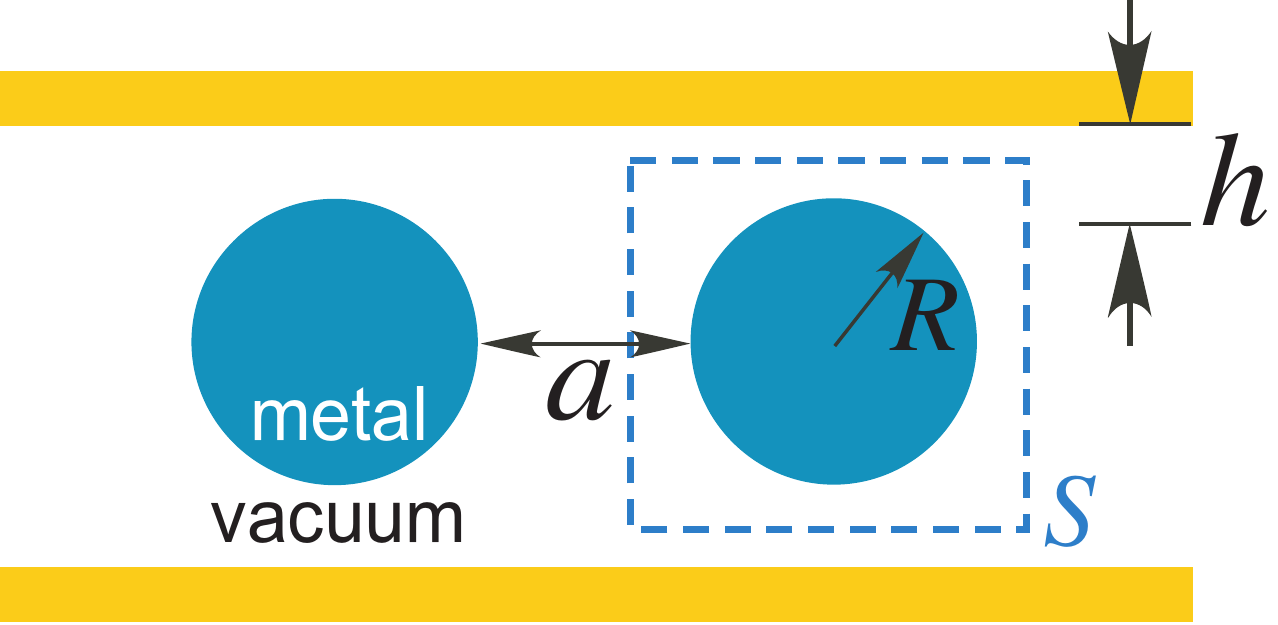}
\centering
\caption{Schematic showing the two-dimensional piston-like
  configuration of~\cite{RahiRo07}: Two perfectly conducting circular
  cylinders of radius $R$, separated by a distance $a$, are sandwiched
  between two perfectly conducting plates (the materials are either
  perfect metallic or perfect magnetic conductors).  The separation
  between the blocks and the cylinder surface is denoted as $h$.}
\label{fig:dblocks}
\end{figure}

In~\citeasnoun{RodriguezMc09:PRA}, we described a method to calculate Casimir forces in the
time domain. Our approach involves a modification of the well-known
stress-tensor method~\cite{Landau:stat2}, in which the force on an
object can be found by integrating the Minkowski stress tensor around
a surface $S$ surrounding the object (~\figref{dblocks}), and over all
frequencies. Our recent approach~\cite{RodriguezMc09:PRA} abandons the
frequency domain altogether in favor of a purely time-domain scheme in
which the force on an object is computed via a series of independent
FDTD calculations in which sources are placed at each point on $S$.
The electromagnetic response to these sources is then integrated in
time against a predetermined function $g(-t)$.  

The main purpose of this approach is to compute the effect of the entire frequency spectrum in a single simulation for each source, rather than a separate set of calculations for each frequency as in most previous work~\cite{Rodriguez07:PRA}.

We exactly transform the problem into a mathematically equivalent
system in which an arbitrary dissipation is introduced.  This
dissipation will cause the electromagnetic response to converge
rapidly, greatly reducing the simulation time.  In particular, a
frequency-independent, spatially uniform conductivity $\sigma$ is
chosen so that the force will converge very rapidly as a function of
simulation time.  For all values of $\sigma$, the force will converge
to the same value, but the optimal $\sigma$ results in the shortest
simulation time and will depend on the system under consideration.
Unless otherwise stated, for the simulations in this paper, we use
$\sigma = 1$ (in units of $2\pi c/a$, $a$ being a typical length scale
in the problem).

In particular, the Casimir force is given by:
\begin{equation}
F_i = \Im \frac{\hbar}{\pi}\int_0^\infty dt ~g(-t) \left(\Gamma^E_i(t) + \Gamma^H_i(t)\right),
\label{eq:time-force}
\end{equation}
where $g(t)$ is a geometry-independent function discussed further in
the Appendix, and the $\Gamma(t)$ are functions of the electromagnetic
fields on the surface $S$ defined in our previous
work~\cite{RodriguezMc09:PRA}.

Written in terms of the electric field response in direction $i$ at
$(t,\vec{x})$ to a source current
$J(t,\vec{x})=\delta(t)\delta(\vec{x}-\vec{x}^\prime)$ in direction
$j$, $E_{ij}(t;\vec{x},\vec{x}^\prime)$, the quantity $\Gamma^E_i(t)$
is defined as:
\begin{eqnarray*}
\Gamma^E_i(t) &\equiv& \int_S dS_j\,(\vec{x}) \left(E_{ij}(t;\vec{x},\vec{x})
- \frac{1}{2}\delta_{ij}\sum_k E_{kk}(t;\vec{x},\vec{x})\right)\\
&\equiv& \int dS_j(\vec{x})\, \Gamma^E_{ij}(t;\vec{x},\vec{x})
\end{eqnarray*}
where $dS_j(\vec{x}) \equiv dS(\vec{x})\, n_j(\vec{x})$, $dS(\vec{x})$
the differential area element and $\vec{n}(\vec{x})$ is the unit
normal vector to $S$ at $\vec{x}$.  A similar definition holds for
$\Gamma^H_i(t)$ involving the magnetic field Green's function
$H_{ij}$.

As described in~\citeasnoun{RodriguezMc09:PRA}, computation of the Casimir force entails
finding both $\Gamma^E(t;\vec{x},\vec{x})$ and the
$\Gamma^H(t;\vec{x},\vec{x})$ field response with a separate
time-domain simulation for every point $\vec{x}\in S$.

While each individual simulation can be performed very efficiently on
modern computers, the surface $S$ will, in general, consist of hundreds
or thousands of pixels or voxels.  This requires a large number of
time-domain simulations, this number being highly dependent upon the
resolution and shape of $S$, making the computation potentially very
costly in practice.

We can dramatically reduce the number of required simulations by
reformulating the force in terms of a harmonic expansion in
$\Gamma^E(t;\vec{x},\vec{x})$, involving the distributed field
responses to distributed currents.  This is done as follows [an
analogous derivation holds for $\Gamma^H(t;\vec{x},\vec{x})$]:

As $S$ is assumed to be a compact surface, we can rewrite
$\Gamma_{ij}(t;\vec{x},\vec{x})$ as an integral over $S$:
\begin{equation}
\Gamma^E_{ij}(t;\vec{x},\vec{x})
= \int_S dS(\vec{x}^\prime)\,\Gamma^E_{ij}(t;\vec{x},\vec{x}^\prime)
\delta_S(\vec{x}-\vec{x}^\prime)
\label{eq:gammaxx}
\end{equation}
where in this integral $dS$ is a scalar unit of area, and $\delta_S$
denotes a $\delta$-function with respect to integrals over the surface
$S$. Given a set of orthonormal basis functions $\lbrace
f_n(\vec{x})\rbrace$ defined on and complete over $S$, we can make the
following expansion of the $\delta$ function, valid for all points
$\vec{x},\vec{x}^\prime \in S$:
\begin{equation}
\delta_S(\vec{x}-\vec{x}^\prime) = \sum_n \bar{f_n}(\vec{x}) f_n(\vec{x}^\prime)
\end{equation}
The $f_n(\vec{x})$ can be an arbitrary set of functions, assuming that
they are complete and orthonormal on $S$~\endnote{Non-orthogonal
  functions may be used, but this case greatly complicates the
  analysis and will not be treated here.}.

\begin{figure}[tb]
\includegraphics[width=0.48\textwidth]{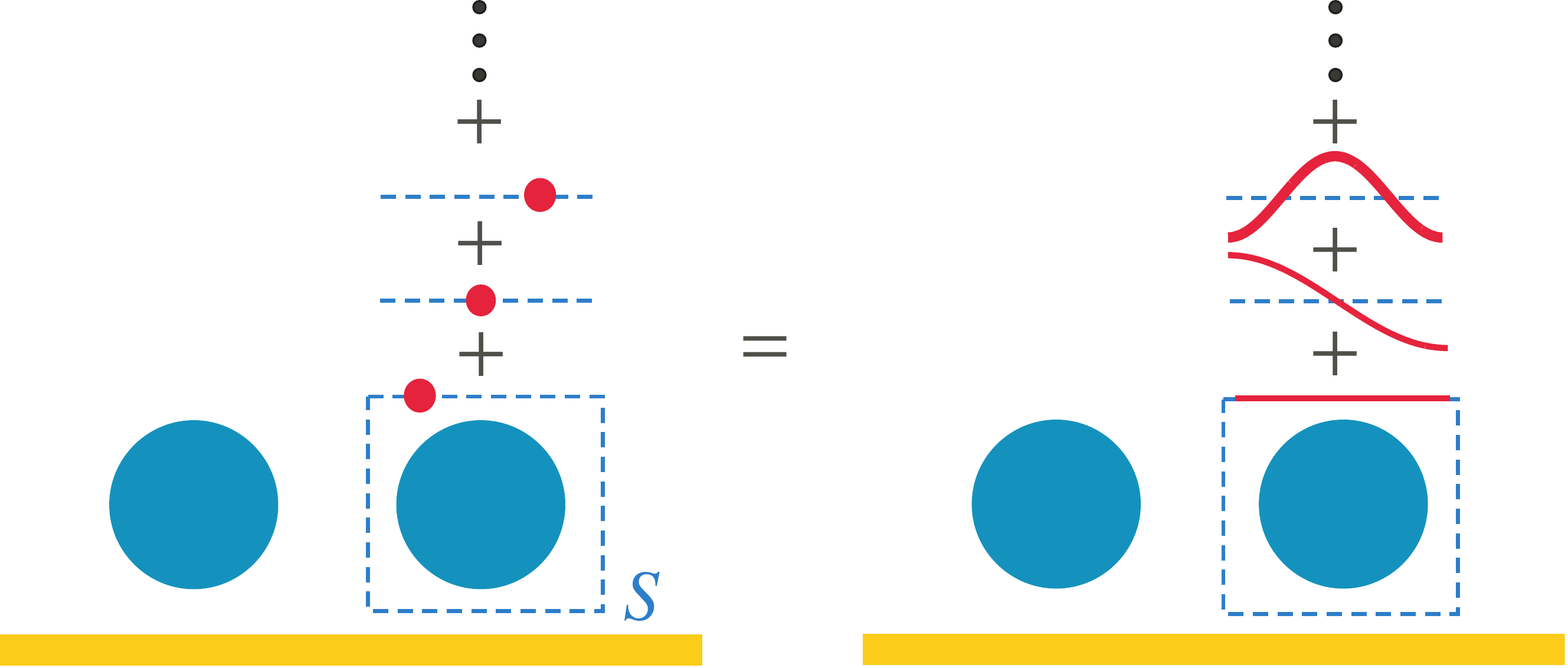}
\centering
\caption{Differing harmonic expansions of the source currents (red) on
  the surface $S$.  The left part shows an expansion using point
  sources, where each dot represents a different simulation.  The
  right part corresponds to using $f_n(\vec{x}) \sim \cos(x)$ for each
  side of $S$.  Either basis forms a complete basis for all functions
  in $S$.}
\label{fig:harmonic}
\end{figure}

Inserting this expansion of the $\delta$-function into~\Eqref{gammaxx}
and rearranging terms yields:
\begin{equation}
\Gamma^E_{ij}(t;\vec{x},\vec{x}) =
\sum_n \bar{f}_n(\vec{x}) \left(\int_S dS(\vec{x}^\prime)\,
 \Gamma^E_{ij}(t;\vec{x},\vec{x}^\prime) f_n(\vec{x}^\prime)\right)
\end{equation}
The term in parentheses can be understood in a physical context: it is
the electric-field response at position $\vec{x}$ and time $t$ to a
current source on the surface $S$ of the form $J(\vec{x},t) =
\delta(t)f_n(\vec{x})$.  We denote this quantity by $\Gamma^E_{ij;n}$:
\begin{equation}
\Gamma^E_{ij;n}(t,\vec{x}) \equiv \int_S dS(\vec{x}^\prime)\,
\Gamma^E_{ij}(t;\vec{x},\vec{x}^\prime)f_n(\vec{x}^\prime)
\end{equation}
where the $n$ subscript indicates that this is a field in response to
a current source determined by
$f_n(\vec{x})$. $\Gamma^E_{ij;n}(t,\vec{x})$ is exactly what can be
measured in an FDTD simulation using a current
$J(\vec{x},t)=\delta(t)f_n(\vec{x})$ for each $n$.  This equivalence
is illustrated in~\figref{harmonic}.

The procedure is now only slightly modified from the one outlined
in~\citeasnoun{RodriguezMc09:PRA}: after defining a geometry and a
surface $S$ of integration, one will additionally need to specify a
set of harmonic basis functions $\lbrace f_n(x)\rbrace$ on $S$.  For
each harmonic moment $n$, one inserts a current function
$\vec{J}(\vec{x},t) = \delta(t)f_n(\vec{x})$ on $S$ and measures the
field response $\Gamma_{ij;n}(\vec{x},t)$.  Summing over all harmonic
moments will yield the total force.

In the following section, we take as our harmonic source basis the
Fourier cosine series for each side of $S$ considered separately,
which provides a convenient and efficient basis for computation.  We
then illustrate its application to systems of perfect conductors and
dielectrics in two and three dimensions.  Three-dimensional systems
with cylindrical symmetry are treated separately, as the harmonic
expansion (as derived in the Appendix) becomes considerably simpler in
this case.

\section{Numerical Implementation}
\label{sec:num-imp}

In principle, any surface $S$ and any harmonic source basis can be
used.  Point sources, as discussed in~\citeasnoun{RodriguezMc09:PRA},
are a simple, although highly inefficient, example.  However, many
common FDTD algorithms (including the one we employ in this paper)
involve simulation on a discretized grid.  For these applications, a
rectangular surface $S$ with an expansion basis separately defined on
each face of $S$ is the simplest.  In this case, the field integration
along each face can be performed to high accuracy and converges
rapidly.  The Fourier cosine series on a discrete grid is essentially
a discrete cosine transform (DCT), a well known discrete orthogonal
basis with rapid convergence properties~\cite{Rao90}.  This in
contrast to discretizing some basis such as spherical harmonics that
are only approximately orthogonal when discretized on a rectangular
grid.

\subsection{Two-dimensional systems}
\label{sec:2d-geoms}

\begin{figure}[tb]
\includegraphics[width=0.48\textwidth]{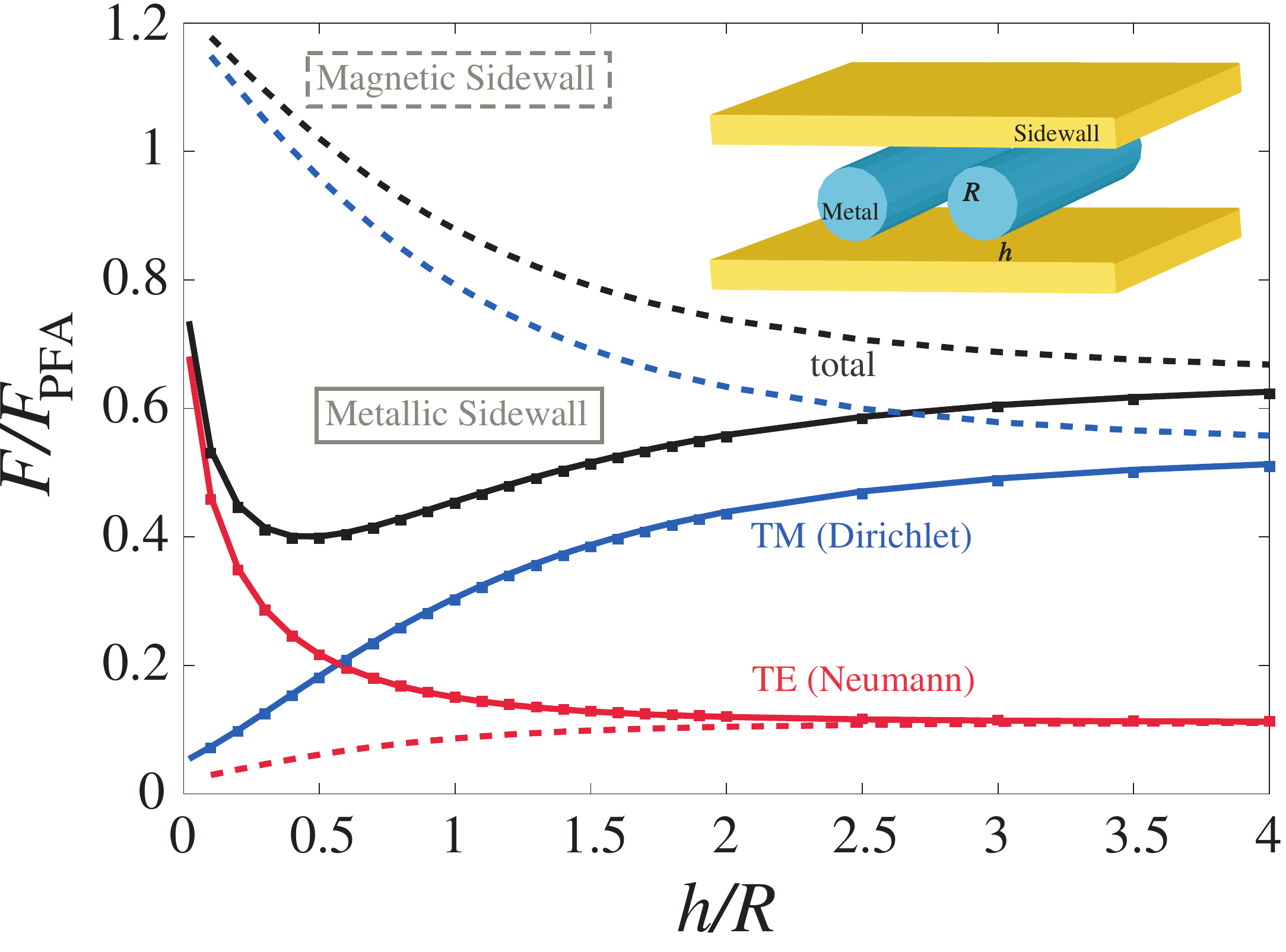}
\centering
\caption{Force for the double cylinders of~\cite{RahiRo07} as a
  function of sidewall separation $h/a$, normalized by the PFA force
  $F_{\mathrm{PFA}}=\hbar c \zeta(3) d/8\pi a^3$.  Red/blue/black
  squares show the TE/TM/total force in the presence of metallic
  sidewalls, as computed by the FDTD method (squares).  The solid
  lines indicate the results from the scattering calculations
  of~\cite{RahiRo07}, showing excellent agreement.  Dashed lines
  indicate the same force components, but in the presence of perfect
  magnetic-conductor sidewalls (computed via FDTD).  Note that the
  total force is nonmonotonic for electric sidewalls and monotonic for
  magnetic sidewalls.}
\label{fig:dblocks-force}
\end{figure}

In this section we consider a variant of the piston-like configuration
of~\citeasnoun{Rodriguez07:PRL}, shown as the inset
to~\figref{dblocks-force}.  This system consists of two cylindrical
rods sandwiched between two sidewalls, and is of interest due to the
non monotonic dependence of the Casimir force between the two blocks
as the vertical wall separation $h/a$ is varied.  The case of perfect
metallic sidewalls $(\varepsilon(x) = -\infty)$ has been solved
previously~\cite{RahiRo07}; here we also treat the case of perfect magnetic conductor
sidewalls $(\mu(x) = -\infty)$ as a simple demonstration of method
using magnetic materials.

While three-dimensional in nature, the system is translation-invariant
in the $z$-direction and involves only perfect metallic or magnetic
conductors.  As discussed in~\citeasnoun{Rodriguez07:PRA} this
situation can actually be treated as the two-dimensional problem
depicted in~\figref{harmonic} using a slightly different form for
$g(-t)$ in~\eqref{time-force} (given in the Appendix).  The reason we
consider the three-dimensional case is that we can directly compare
the results for the case of metallic sidewalls to the high-precision
scattering calculations of~\citeasnoun{RahiRo07} (which uses a
specialized exponentially convergent basis for cylinder/plane
geometries).

For this system, the surface $S$ consists of four faces, each of which
is a line segment of some length $L$ parametrized by a single variable
$x$.  We employ a cosine basis for our harmonic expansion on each face
of $S$.  The basis functions for each side are then:

\begin{equation}
f_n(x) = \sqrt{\frac{2}{L}}\cos\left(\frac{n\pi x}{L}\right),~n=0,1,\ldots
\end{equation}
where $L$ is the length of the edge, and $f_n(x) = 0$ for
all points $x$ not on that edge of $S$.  These functions, and their
equivalence to a computation using $\delta$-function sources as basis
functions, are shown in~\figref{harmonic}.

In the case of our FDTD algorithm, space is discretized on a Yee
grid~\cite{Taflove00}, and in most cases $x$ will turn out to lie in
between two grid points.  One can run separate simulations in which
each edge of $S$ is displaced in the appropriate direction so that all
of its sources lie on a grid point.  However, we find that it is
sufficient to place suitably averaged currents on neighboring grid
points, as several available FDTD implementations provide features to
accurately interpolate currents from any location onto the grid.

The force, as a function of the vertical sidewall separation $h/a$,
and for both TE and TM field components, is shown
in~\figref{dblocks-force} and checked against previously known results
for the case of perfect metallic sidewalls~\cite{RahiRo07}.  We also
show the force (dashed lines) for the case of perfect magnetic
conductor sidewalls.

In the case of metallic sidewalls, the force is nonmonotonic in $h/a$.
As explained in~\citeasnoun{RahiRo07}, this is due to the competition
between the TM force, which dominates for large $h/a$ but is
suppressed for small $h/a$, and the TE force, which has the opposite
behavior, explained via the method of images for the conducting walls.
Switching to perfect magnetic conductor sidewalls causes the TM force
to be enhanced for small $h/a$ and the TE force to be suppressed,
because the image currents flip sign for magnetic conductors compared
to electric conductors.  As shown in~\figref{dblocks-force}, this
results in a monotonic force for this case.

\begin{figure}[tb]
\includegraphics[width=0.48\textwidth]{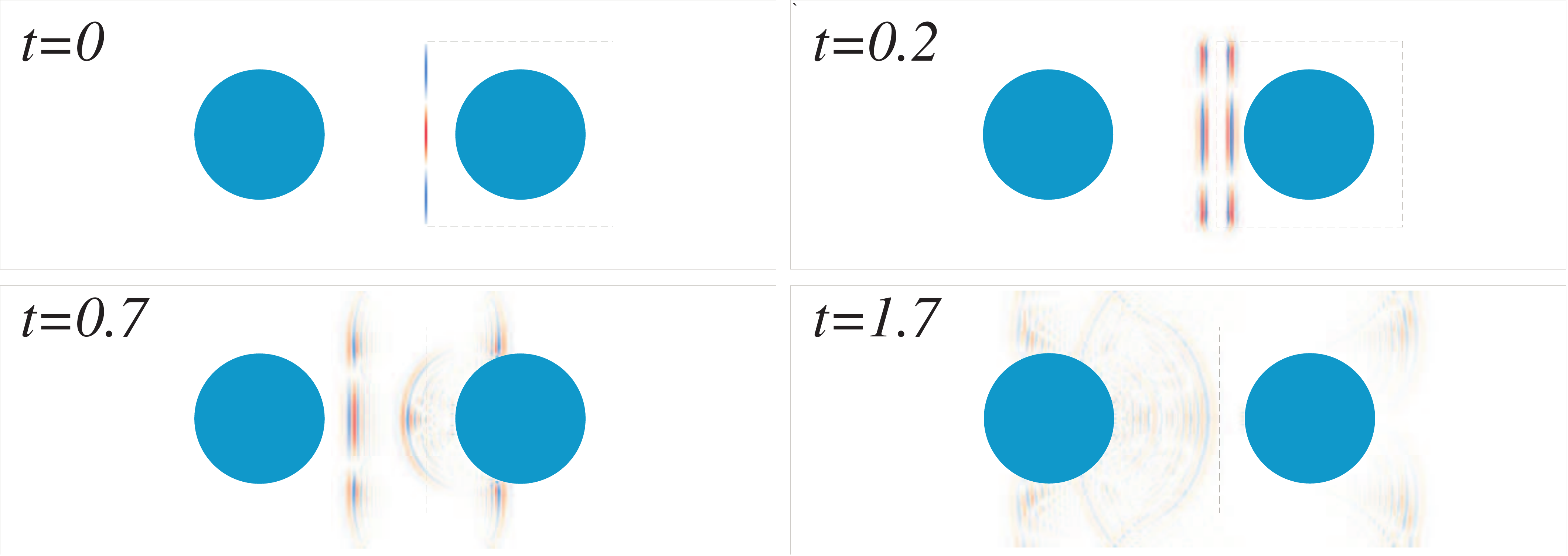}
\centering
\caption{$\Gamma^E_{yy;n=2}(t,\vec{x})$ snapshots (blue/white/red =
  positive/zero/negative) for the $n=2$ term in the harmonic cosine
  expansion on the leftmost face of $S$ for the double blocks
  configuration of~\cite{Rodriguez07:PRL} at selected times (in units
  of $a/c$).}
\label{fig:visualization}
\end{figure}

The result of the above calculation is a time-dependent field similar
to that of~\figref{visualization}, which when manipulated as
prescribed in the previous section, will yield the Casimir force.  As
in~\citeasnoun{RodriguezMc09:PRA}, our ability to express
the force for a dissipationless system (perfect-metal blocks in
vacuum) in terms of the response of an artificial dissipative system
$(\sigma \neq 0)$ means that the fields, such as those shown
in~\figref{visualization}, rapidly decay away, and hence only a short
simulation is required for each source term.  

\begin{figure}[tb]
\includegraphics[width=0.48\textwidth]{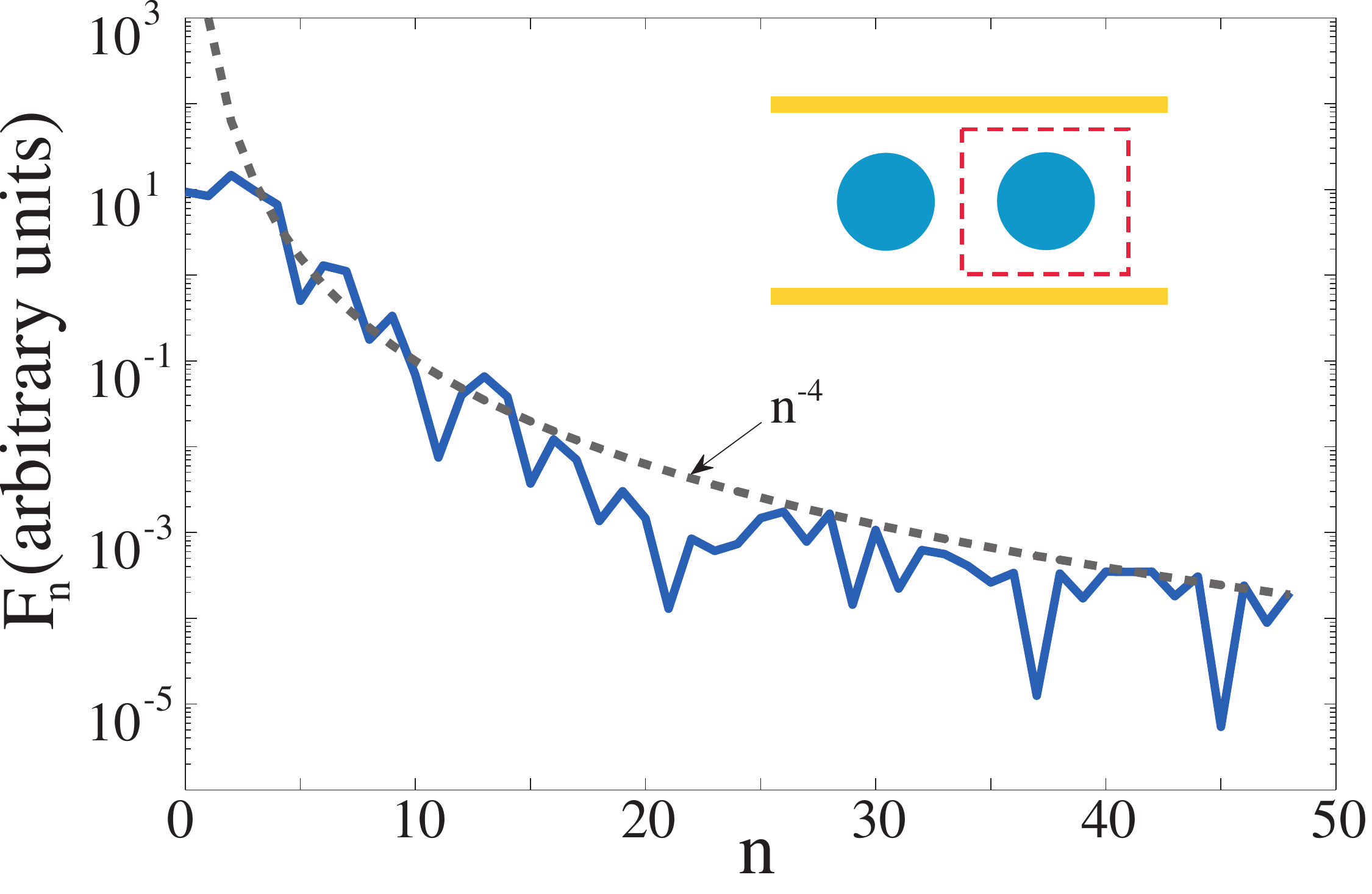}
\centering
\caption{Relative contribution of harmonic moment $n$ in the cosine
  basis to the total Casimir force for the double blocks configuration
  (shown in the inset)}
\label{fig:harmonic-convergence}
\end{figure}

In addition,~\figref{harmonic-convergence} shows the convergence of
the harmonic expansion as a function of $n$.  Asymptotically for large
$n$, an $n^{-1/4}$ power law is clearly discernible.  The explanation
for this convergence follows readily from the geometry of $S$: the
electric field $\vec{E}(\vec{x})$, when viewed as a function along
$S$, will have nonzero first derivatives at the corners.  However, the
cosine series used here always has a vanishing derivative.  This
implies that its cosine transform components will decay asymptotically
as $n^{-2}$~\cite{boyd01:book}.  As $\Gamma^E$ is related to the
correlation function $\langle
\vec{E}(\vec{x})\vec{E}(\vec{x})\rangle$, their contributions will
decay as $n^{-4}$. One could instead consider a Fourier series defined
around the whole perimeter of $S$, but the convergence rate will be
the same because the derivatives of the fields will be discontinuous
around the corners of $S$.  A circular surface would have no corners
in the continuous case, but on a discretized grid would effectively
have many corners and hence poor convergence with resolution.

\subsection{Dispersive Materials}
Dispersion in FDTD in general requires fitting an actual dispersion to
a simple model (eg. a series of Lorentzians or Drude peaks).  Assuming
this has been done, these models can then be analytically continued
onto the complex conductivity contour.

As an example of a calculation involving dispersive materials, we
consider in this section a geometry recently used to measure the
classical optical force between two suspended
waveguides~\cite{Li09:arxiv}, confirming a
prediction~\cite{Povinelli05} that the sign of the classical force
depends on the relative phase of modes excited in the two waveguides.
We now compute the Casimir force in the same geometry, which consists
of two identical silicon waveguides in empty space.  We model silicon
as a dielectric with dispersion given by:
\begin{equation}
\varepsilon(\omega) = \varepsilon_f + \frac{\varepsilon_f - \varepsilon_0}{1 - \left(\frac{\omega}{\omega_0}\right)^2}
\label{eq:silicon}
\end{equation}
where $\omega_0 = 6.6\times 10^{15}$ rad/sec, and $\varepsilon_0 =
1.035$, $\varepsilon_f = 11.87$. This dispersion can be implemented in
FDTD by the standard technique of auxiliary differential
equations~\cite{Taflove00} mapped into the complex-$\omega$ plane as
explained in~\citeasnoun{RodriguezMc09:PRA}.

\begin{figure}[tb]
\includegraphics[width=0.48\textwidth]{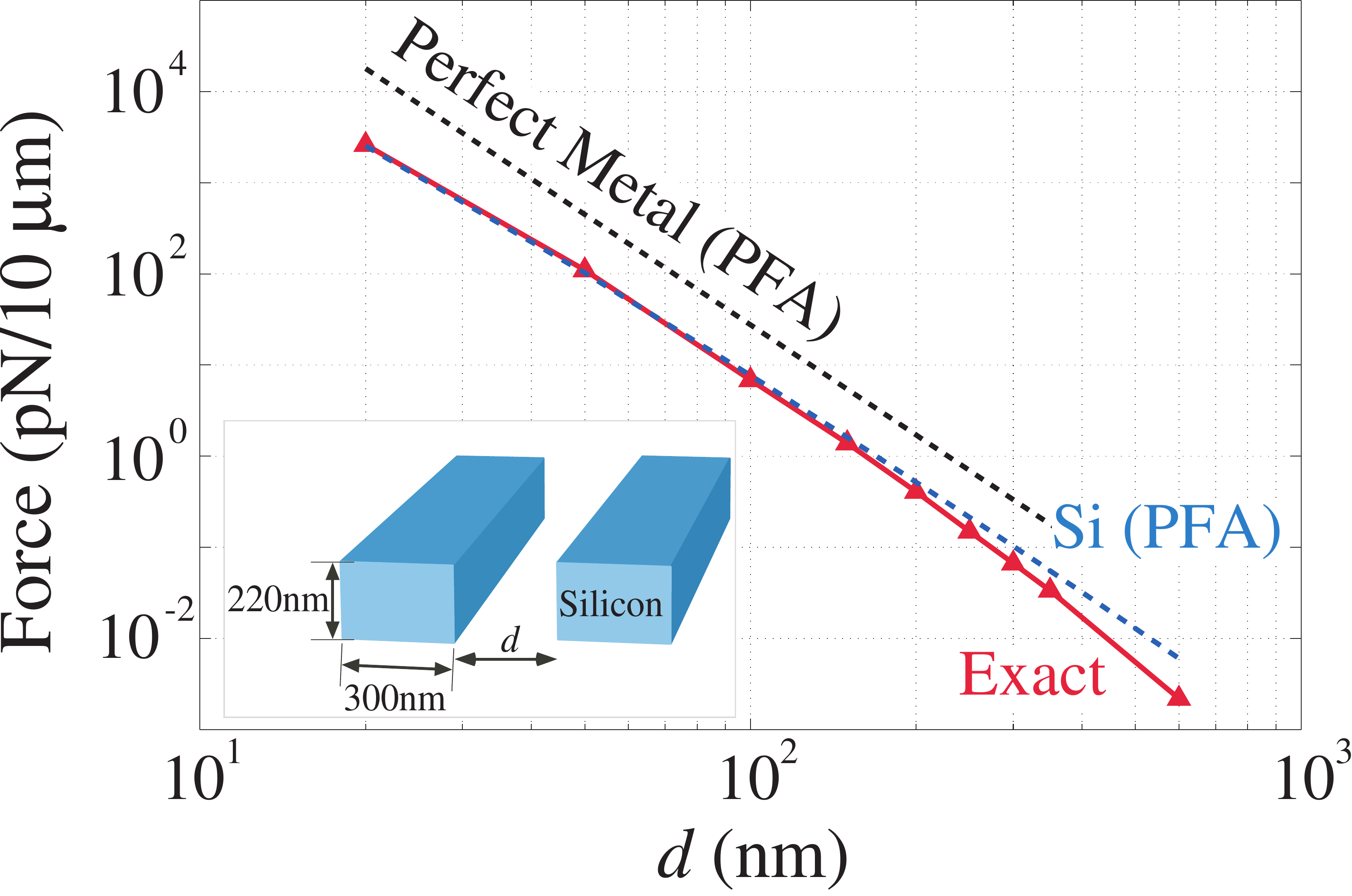}
\centering
\caption{Force per unit length between long silicon waveguides
  suspended in air~\cite{Li09:arxiv}, determined by the FDTD method
  (red triangles). Also shown is the analogous one-dimensional
  computation assuming silicon plates of finite thickness (blue
  dashes), and the result $F = \pi^2/240d^4 A$ for perfect metals
  (assuming plate area $A$ equal to the interaction area of the
  waveguides).}
\label{fig:tang}
\end{figure}

The system is translation-invariant in the $z$ direction.  If it
consisted only of perfect conductors, we could use the trick of the
previous section and compute the force in only one 2d simulation.
However, dielectrics hybridize the two polarizations and require an
explicit $k_z$ integral, as discussed in~\citeasnoun{Rodriguez07:PRA}.
Each value of $k_z$ corresponds to a separate two-dimensional
simulation with Bloch-periodic boundary conditions.  The value of the
force for each $k_z$ is smooth and rapidly decaying, so in general
only a few $k_z$ points are needed.

To simulate the infinite open space around the waveguides, it is ideal
to have ``absorbing boundaries'' so that waves from sources on $S$ do
no reflect back from the boundaries.  We employ the standard technique
of perfectly matched layers (PML), which are a thin layer of
artificial absorbing material placed adjacent to the boundary and
designed to have nearly zero reflections~\cite{Taflove00}.  The
results are shown in red in~\figref{tang}.  We also show (in blue) the
force obtained using the proximity force approximation (PFA)
calculations based on the Lifshitz formula~\cite{Lifshitz55,
  Dzyaloshinskii61}.  For the PFA, we assume two parallel silicon
plates, infinite in both directions perpendicular to the force and
having the same thickness as the waveguides in the direction parallel
to the force, computing the PFA contribution from the surface area of
the waveguide.  As expected, at distances smaller than the waveguide
width, the actual and PFA results are in good agreement, while as the
waveguide separation increases, the PFA becomes more inaccurate.  For
example, by a separation of 300nm, the PFA result is off by 50$\%$.
We also show for comparison the force for the same surface between two
perfectly metallic plates, also assuming infinite extent in both
transverse directions.

\subsection{Three dimensions with cylindrical symmetry}
\label{sec:3d-cyls}

In the case of cylindrical symmetry, we can employ a cylindrical
surface $S$ and a complex exponential basis $e^{im\phi}$ in the $\phi$
direction.  For a geometry with cylindrical symmetry and a separable
source with $e^{im\phi}$ dependence, the resulting fields are also
separable with the same $\phi$ dependence, and the unknowns reduce to
a two-dimensional $(r,z)$ problem for each $m$.  This results in a
substantial reduction in computational costs compared to a full
three-dimensional computation.

Treating the reduced system as a two-dimensional space with
coordinates $(r,z)$, the expression for the force (as derived in the
Appendix) is now:
\begin{equation}
F_i = \sum_{n} \int_0^\infty dt\,
\Im[g(-t)]\int_S ds_j(\vec{x})\,\Gamma_{ij;n}(\vec{x},t)
\label{eq:force-cyl}
\end{equation}
where the $m$-dependence has been absorbed into the definition of
$\Gamma$ above:
\begin{equation}
\Gamma_{ij;n}(x,t) \equiv \Gamma_{ij;n,m=0}(\vec{x},t) 
+ 2\sum_{m>0}\Re[\Gamma_{ij;n,m}(\vec{x},t)],
\label{eq:gamma-cyl}
\end{equation}
and $ds_j = ds\,n_j(\vec{x})$, $ds$ being a one-dimensional Cartesian
line element.  As derived in the Appendix, the Jacobian factor $r$
obtained from converting to cylindrical coordinates cancels out, so
that the one-dimensional ($r$-independent) measure $ds$ is the
appropriate one to use in the surface integration.  Also, the
$2\Re[\cdots]$ comes from the fact that the $+m$ and $-m$ terms are
complex conjugates.  Although the exponentials $e^{im\phi}$ are
complex, only the real part of the field response appears
in~\eqref{gamma-cyl}, allowing us to use $\Im[g(-t)]$ alone
in~\eqref{force-cyl}.

Given an $e^{im\phi}$ dependence in the fields, one can write
Maxwell's equations in cylindrical coordinates to obtain a
two-dimensional equation involving only the fields in the $(r,z)$
plane.  This simplification is incorporated into many FDTD solvers, as
in the one we currently employ~\cite{Farjadpour06}, with the
computational cell being restricted to the $(r,z)$ plane and $m$
appearing as a parameter.  When this is the case, the implementation
of cylindrical symmetry is almost identical to the two-dimensional
situation.  The only difference is that now there is an additional
index $m$ over which the force must be summed.

\begin{figure}[tb]
\includegraphics[width=0.48\textwidth]{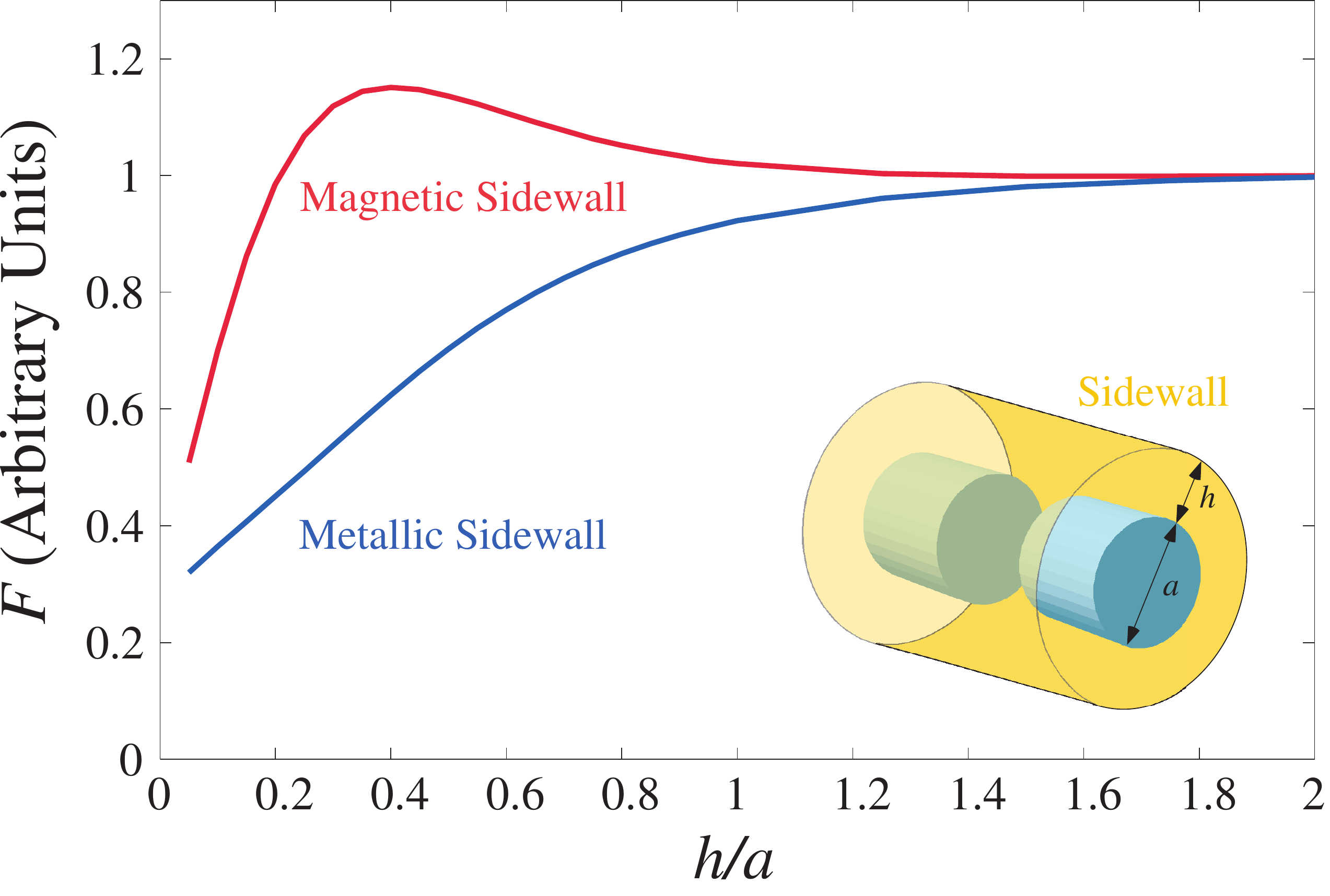}
\centering
\caption{Force as a function of outer sidewall spacing $h/a$ for the
  cylindrically-symmetric piston configuration shown in the figure.
  Both plates are perfect metals, and the forces for both perfect
  metallic and perfect magnetic conductor sidewalls are shown.  Note that in
  contrast to~\figref{dblocks-force}, here the force is monotonic in
  $h/a$ for the metallic case and non monotonic for the magnetic case.}
\label{fig:3d-blocks-force}
\end{figure}

To illustrate the use of this algorithm with cylindrical symmetry, we
examine the 3d system shown in the inset of~\figref{3d-blocks-force}.
This configuration is similar to the configuration of cylindrical rods
of~\figref{dblocks-force}, except that instead of translational ($z$)
invariance we instead impose rotational ($\phi$) invariance. In this
case, the two sidewalls are joined to form a cylindrical tube.  We
examine the force between the two blocks as a function of $h/a$ (the
$h=0$ case has been solved analytically~\cite{Marachevsky07}).

Due to the two-dimensional nature of this problem, computation time is
comparable to that of the two-dimensional double block geometry of the
previous section.  Rough results (at resolution 40, accurate to within
a few percentage points) can be obtained rapidly on a single computer
(about 5 minutes running on 8 processors) are shown
in~\figref{3d-blocks-force} for each value of $h/a$.  Only indices
$n,m\in\lbrace 0,1,2\rbrace$ are needed for the result to have
converged to within 1$\%$, after which the error is dominated by the
spatial discretization.  PML is used along the top and bottom walls
of the tube.

In contrast to the case of two pistons with translational symmetry,
the force for metallic sidewalls is monotonic in $h/a$.  Somewhat
surprisingly, when the sidewalls are switched to perfect magnetic
conductors the force becomes non monotonic again.  Although the use of
perfectly magnetic conductor sidewalls in this example is unphysical, it
demonstrates the use of a general-purpose algorithm to examine the
material-dependence of the Casimir force.  If we wished to use
dispersive and/or anisotropic materials, no additional code would be
required.

\subsection{Periodic Boundary Conditions}

\begin{figure}[tb]
\includegraphics[width=0.48\textwidth]{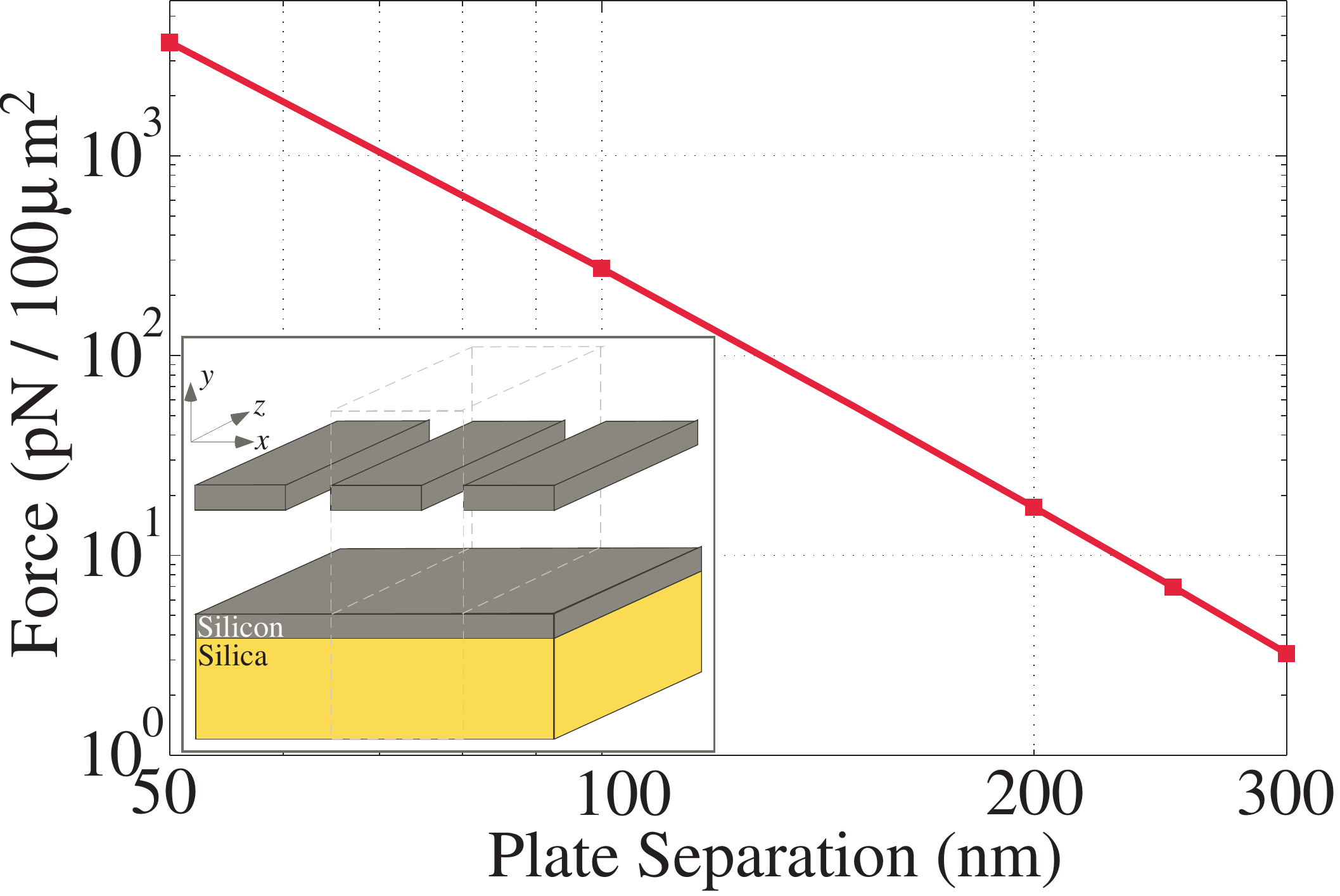}
\centering
\caption{The Casimir force between a periodic array of Silicon
  waveguides and a Silicon/Silica substrate, as the array/substrate
  separation is varied.  The system is periodic in the $x$-direction
  and translation-invariant in the $z$-direction, so the computation
  involves a set of two-dimensional simulations.}
\label{fig:grating}
\end{figure}

Periodic dielectric systems are of interest in many applications.  The
purpose of this section is to demonstrate computations involving a
periodic array of dispersive silicon dielectric waveguides above a
silica substrate, shown in~\figref{grating}.

As discussed in~\citeasnoun{Rodriguez07:PRA}, the Casimir force for
periodic systems can be computed as in integral over all Bloch
wavevectors in the directions of periodicity.  Here, as there are two
directions, $x$ and $z$, that are periodic (the latter being the limit
in which the period goes to zero).  The force is then given by:

\begin{equation}
\int_0^\infty \int_0^\infty F_{k_z,k_x}dk_z dk_x
\end{equation}

where $F_{k_z,k_x}$ is the force computed from one simulation of the
unit cell using Bloch-periodic boundary conditions with wavevector
$\vec{k} = (k_x,0,k_z)$.  In the present case, the unit cell is of
period 1$\mu$m in the $x$ direction and of zero length in the $z$
direction, so the computations are effectively two-dimensional
(although they must be integrated over $k_z$).

We use the dispersive model of~\eqref{silicon} for silicon, while for
silica we use~\cite{Rodriguez07:PRL}
\begin{equation}
\epsilon(\omega) = 1 + \sum_{j=1}^3 \frac{C_j \omega_j^2}{\omega_j^2 - \omega^2}
\end{equation}
where $(C_1,C_2,C_3)=(0.829,0.095,1.098)$ and
$(\omega_1,\omega_2,\omega_3)=(0.867,1.508,203.4)\times 10^{14}$
(rad/sec). 

%As discussed in~\cite{Rodriguez07:PRA}, although the physical system
%is unbounded in three dimensions, the two dimensions of periodicity
%(one with finite period $a$ and the other obtained from the limit that
%the period goes to zero) allow the stress tensor to be decomposed into
%Bloch modes, indexed in this case by the wavevectors $k_y$ and $k_z$.
%The force on an object (in the $x$-direction) is then obtained by
%inserting an infinite surface (as shown)...

\subsection{Full 3d computations}
\label{sec:full-3d}

\begin{figure}[tb]
\includegraphics[width=0.42\textwidth]{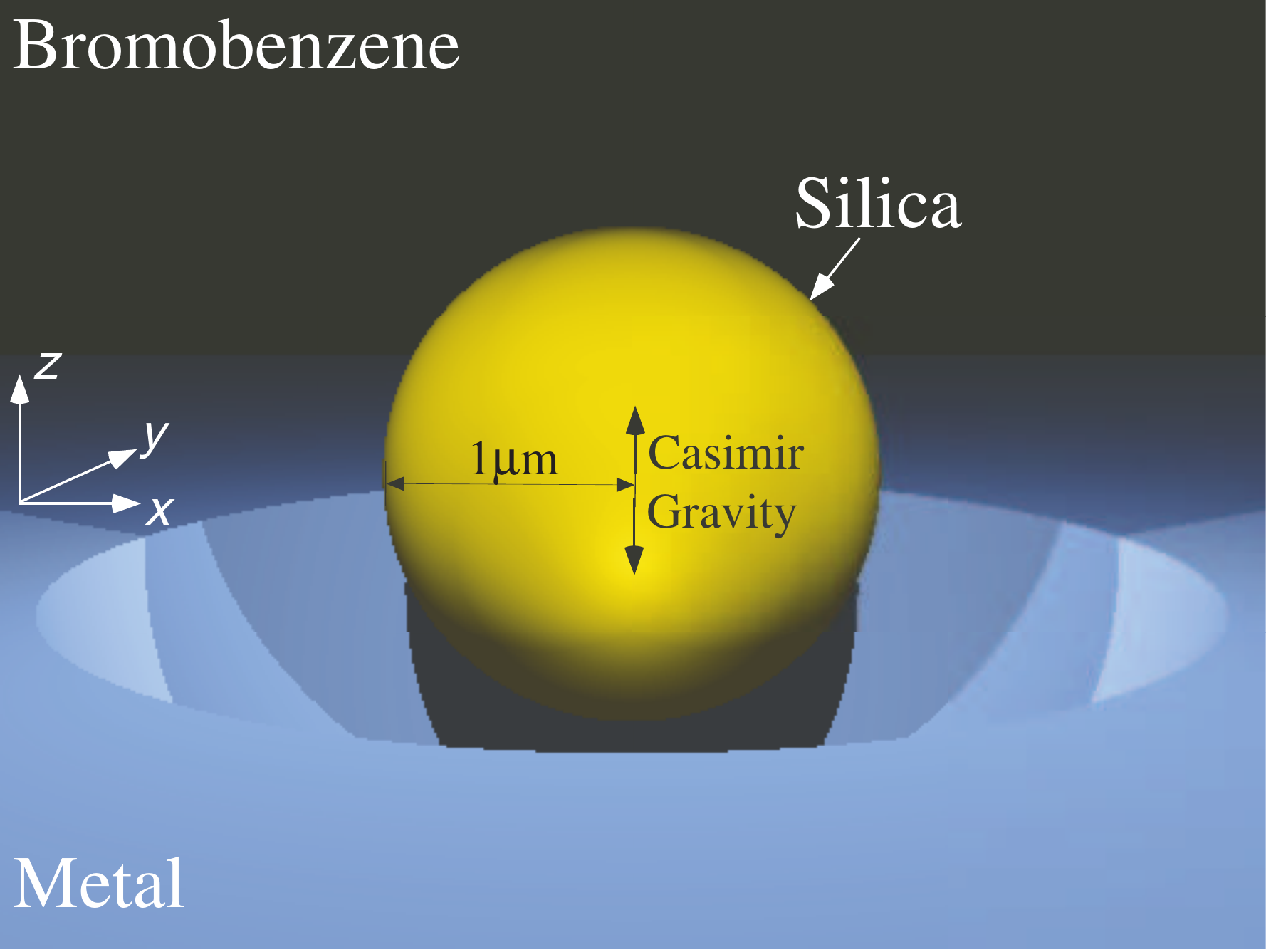}
\centering
\caption{Three-dimensional configuration showing stable levitation.
  At the equilibrium point, the force of gravity counters the Casimir
  force, while the Casimir force from the walls of the spherical
  indentation confine the sphere laterally}
\label{fig:3d-indented-sphere}
\end{figure}

As a final demonstration, we compute the Casimir force for a fully
three-dimensional system, without the use of special symmetries.  The
system used is depicted in~\figref{3d-indented-sphere}.

This setup demonstrates stable levitation with the aid of a fluid
medium, which has been explored previously
in~\citeasnoun{Rodriguez08:PRL}.  With this example, we present a
setup similar to that used previously to measure repulsive Casimir
forces~\cite{Munday09}, with the hope that this system may be
experimentally feasible.

A silica sphere sits atop a perfect metal plane which has a spherical
indentation in it.  The sphere is immersed in bromobenzene.  As the
system satisfies $\varepsilon_{\mathrm{sphere}} < \varepsilon_{\mathrm{fluid}} <
  \varepsilon_{\mathrm{plane}}$, the sphere feels a repulsive Casimir force
  upwards~\cite{Munday09}.  This is balanced by the downward force of gravity, which
  confines the sphere vertically.  In addition, the Casimir repulsion
  from the sides of the spherical indentation confine the sphere in
  the lateral direction.  The radius of the sphere is $1\mu$m, and the
  circular indentation in the metal is formed from a circle of radius
  $2\mu$m, with a center $1\mu$m above the plane.  For computational
  simplicity, in this model we neglect dispersion and use the
  zero-frequency values for the dielectrics, as the basic effect does
  not depend upon the dispersion (the precise values for the
  equilibrium separations will be changed with dispersive materials).
  These are $\varepsilon=2.02$ for silica and $\varepsilon=4.30$.

\begin{figure}[tb]
\includegraphics[width=0.48\textwidth]{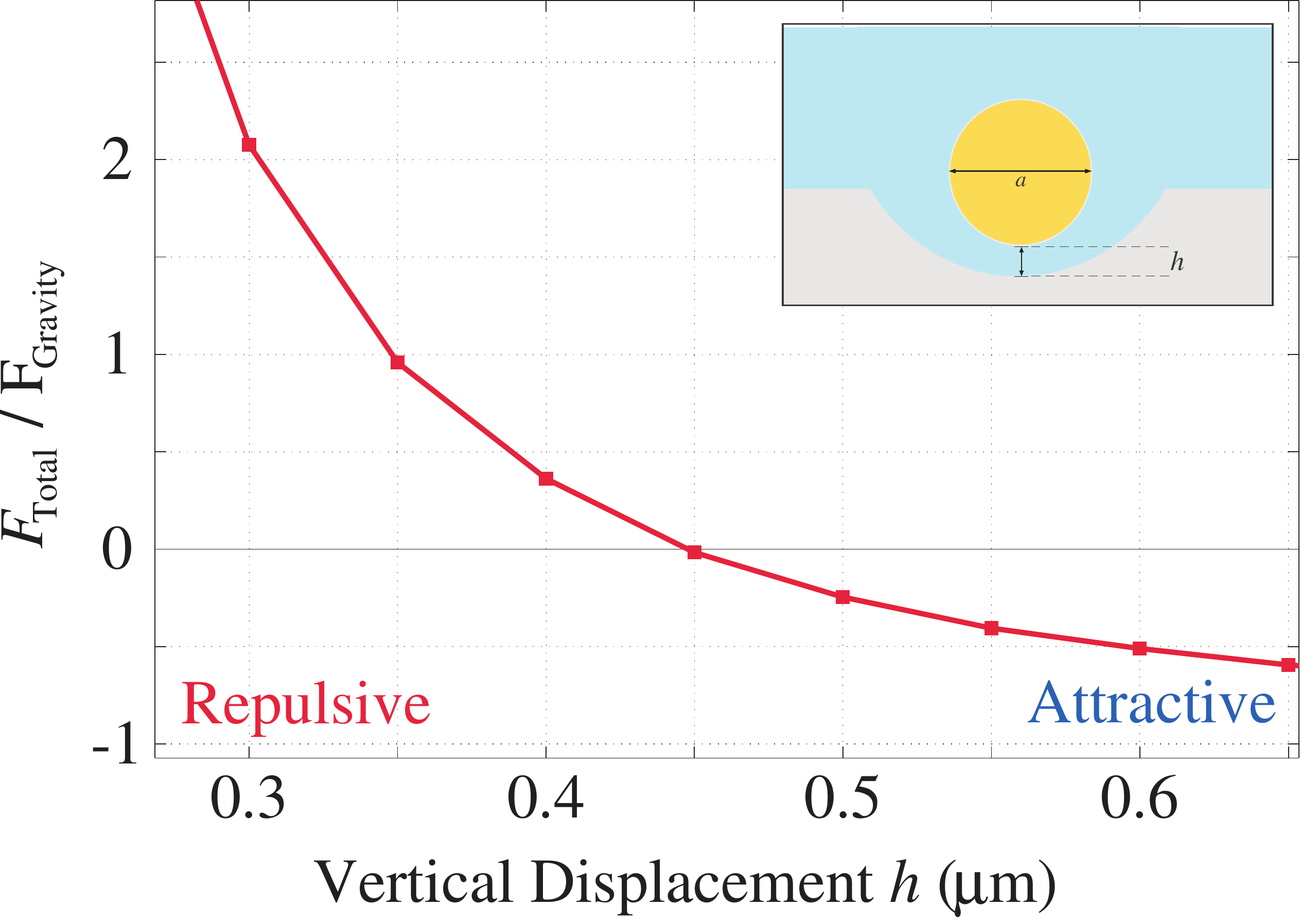}
\centering
\caption{Total (Casimir + gravity) vertical ($z$) force on the silica
  sphere (depicted in the inset) as the height $h$ of the sphere's
  surface above the indentation surface is varied.  The point of
  vertical equilibrium occurs at $h\sim 450$ nm.}
\label{fig:3d-z-force}
\end{figure}

An efficient strategy to determine the stable point is to first
calculate the force on the glass sphere when its axis is aligned with
the symmetry axis of the indentation.  This configuration is
cylindrically-symmetric and can be efficiently computed as in the
previous section.  Results for a specific configuration, with a
sphere radius of $500$ nm and an indentation radius of $1~\mu$m, are
shown in~\figref{3d-z-force}.  

\begin{figure}[tb]
\includegraphics[width=0.49\textwidth]{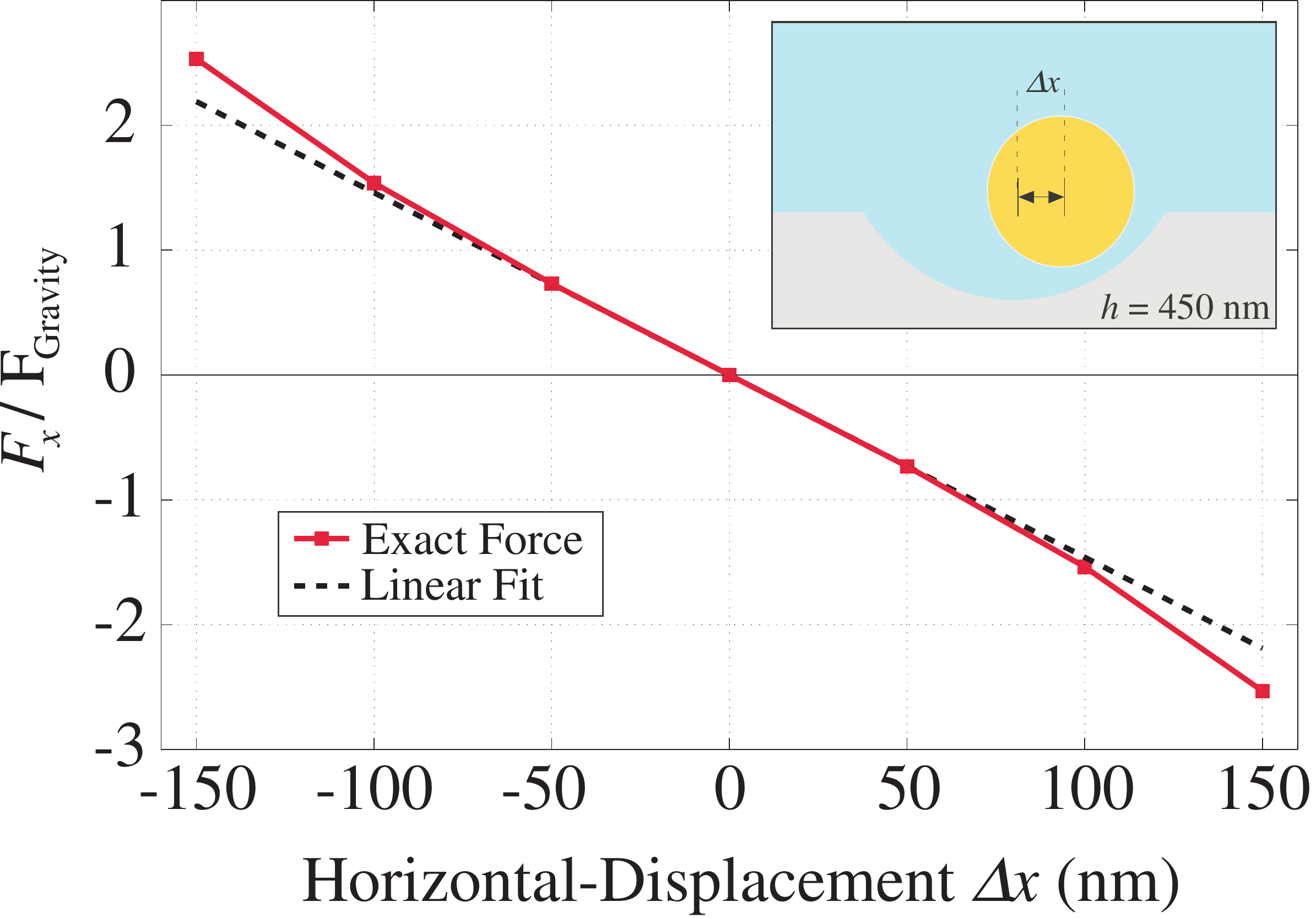}
\centering
\caption{Casimir restoring force on the sphere as a function of
  lateral displacement $dx$, when the vertical position is fixed at
  $h=450$ nm, the height at which gravity balances the Casimir force}.
\label{fig:3d-x-force}
\end{figure}

The force of gravity is balanced against the Casimir force at a height
of $h=450$ nm.  To determine the strength of lateral confinement, we
perform a fully three-dimensional computation in which the center of
the sphere is displaced laterally from equilibrium by a distance $dx$
(the vertical position is held fixed at the equilibrium value $h =
450$nm).  The results are shown in~\figref{3d-x-force}.  It is seen
that over a fairly wide range ($|\Delta x| < 100$ nm) the linear term is a
good approximation to the force, whereas for larger displacements the
Casimir force begins to increase more rapidly.  Of course, at these
larger separations the vertical force is no longer zero, due to the
curvature of the indentation, and so must be re-computed as well.

The fully three-dimensional computations are rather large, and require
roughly a hundred CPU hours per force point.  However, these Casimir
calculations parallelize very easily -- every source term,
polarization, and $k$-point can be computed in parallel, and
individual FDTD calculations can be parallelized in our existing
software -- so we can compute each force point in under an hour on a
supercomputer (with 1000+ processors).  In contrast, the 2d and
cylindrical calculations require tens of minutes per force point.  We
believe that this method is usable in situations involving complex
three-dimensional materials (e.g, periodic systems or systems with
anisotropic materials).

\section{Concluding Remarks}

We have demonstrated a practical implementation of a general FDTD
method for computing Casimir forces via a harmonic expansion in source
currents.  The utility of such a method is that many different systems
(dispersive, anisotropic, periodic boundary conditions) can all be
simulated with the same algorithm.

In practice, the harmonic expansion converges rapidly with higher
harmonic moments, making the overall computation complexity of the
FDTD method $O( N^{1+1/d})$ for $N$ grid points and $d$ spatial
dimensions.  This arises from the $O(N)$ number of computations needed
for one FDTD time step, while the time increment used will vary
inversely with the spatial resolution~\cite{Taflove00}, leading to
$O(N^{1/d})$ time steps per simulation.  In addition, there is a
constant factor proportional to the number of terms retained in the
harmonic expansion, as an independent simulation is required for each
term.  For comparison, without a harmonic expansion one would have to
run a separate simulation for each point on $S$.  In that case, there
would be $O(N^{(d-1)/d})$ points, leading to an overall computational
cost of $O(N^2)$~\cite{RodriguezMc09:PRA}.

We do not claim that this is the most efficient technique for
computing Casimir forces, as there are other works that have also
demonstrated very efficient methods capable of handling arbitrary
three-dimensional geometries, such as a recently-developed
boundary-element method~\cite{ReidRo09}.  However, these
integral-equation methods and their implementations must be
substantially revised when new types of materials or boundary
conditions are desired that change the underlying Green's function
(e.g., going from metals to dielectrics, periodic boundary conditions,
or isotropic to anisotropic materials), whereas very general FDTD
codes, requiring no modifications, are available off-the-shelf.

\section{Acknowledgments}

We are grateful to S. Jamal Rahi for sharing his scattering algorithm
with us.  We are also grateful to Peter Bermel and Ardavan Oskooi for
helpful discussions.

\section{Appendix}
\label{sec:app}

\subsection{Simplified computation of $g(t)$}
\label{sec:new-gt}

In~\citeasnoun{RodriguezMc09:PRA} we introduced a geometry-independent
function $g(t)$, which resulted from the Fourier transform of a
certain function of frequency, termed $g(\xi)$, which is given
by~\cite{RodriguezMc09:PRA}:
\begin{equation}
g(\xi) = -i\xi\left(1+\frac{i\sigma}{\xi}\right)
\frac{1+i\sigma/2\xi}{\sqrt{1+i\sigma/\xi}}\Theta(\xi)
\end{equation}
Once $g(t)$ is known, it can be integrated against the fields in
time, allowing one to compute a decaying time-series which will,
when integrated over time, yield the correct Casimir force.

$g(\xi)$ has the behavior that it diverges in the high-frequency
limit.  For large $\xi$, $g(\xi)$ has the form:
\begin{equation}
g(\xi) \rightarrow g_{1}(\xi) \equiv \frac{\xi}{i}\Theta(\xi) + \sigma \Theta(\xi)
~~\mathrm{as}~\xi\rightarrow\infty
\label{eq:g-large}
\end{equation}
Viewing $g_1(\xi)$ as a function, we could only compute its Fourier
transform $g_1(t)$ by introducing a cutoff in the frequency integral
at the Nyquist frequency, since the time signal is only defined up to
a finite sampling rate and the integral of a divergent function may
appear to be undefined in the limit of no cutoff.

Applying this procedure to compute $g(-t)$ yields a time series that
has strong oscillations at the Nyquist frequency.  The amplitude of
these oscillations can be quite high, increasing the time needed to
obtain convergence and also making any physical interpretation of the
time series more difficult.

These oscillations are entirely due to the high-frequency behavior of
$g(\xi)$, where $g(\xi) \sim g_1(\xi)$.  However, $g(t)$ and $g(\xi)$
only appear when when they are being integrated against smooth,
rapidly decaying field functions $\Gamma(\vec{x},t)$ or
$\Gamma(\vec{x},\xi)$.  In this case, $g$ can be viewed as a tempered
distribution (such as the $\delta$-function)~\cite{bigrudin}.
Although $g(\xi)$ diverges for large $\xi$, this divergence is only a
power law, so it is a tempered distribution and its Fourier transform
is well-defined without any truncation. In particular, the Fourier
transform of $g_1(\xi)$ is given by:
\begin{equation}
g_1(-t) = \frac{i}{2\pi}\left(\frac{1}{t^2} + \frac{\sigma}{t}\right)
\end{equation}

Adding and subtracting the term $g_1(\xi)$ from $g_(\xi)$, the
remaining term decays to zero for large $\xi$ and can be Fourier
transformed numerically without the use of a high-frequency cutoff,
allowing $g(-t)$ to be computed as the sum of $g_1(t)$ plus the
Fourier transform of a well-behaved function.  This results in a much
smoother $g(-t)$ which will give the same final force as the $g(-t)$
used in~\citeasnoun{RodriguezMc09:PRA}, but will also have a much more
well-behaved time dependence.

\begin{figure}
\includegraphics[width=0.48\textwidth]{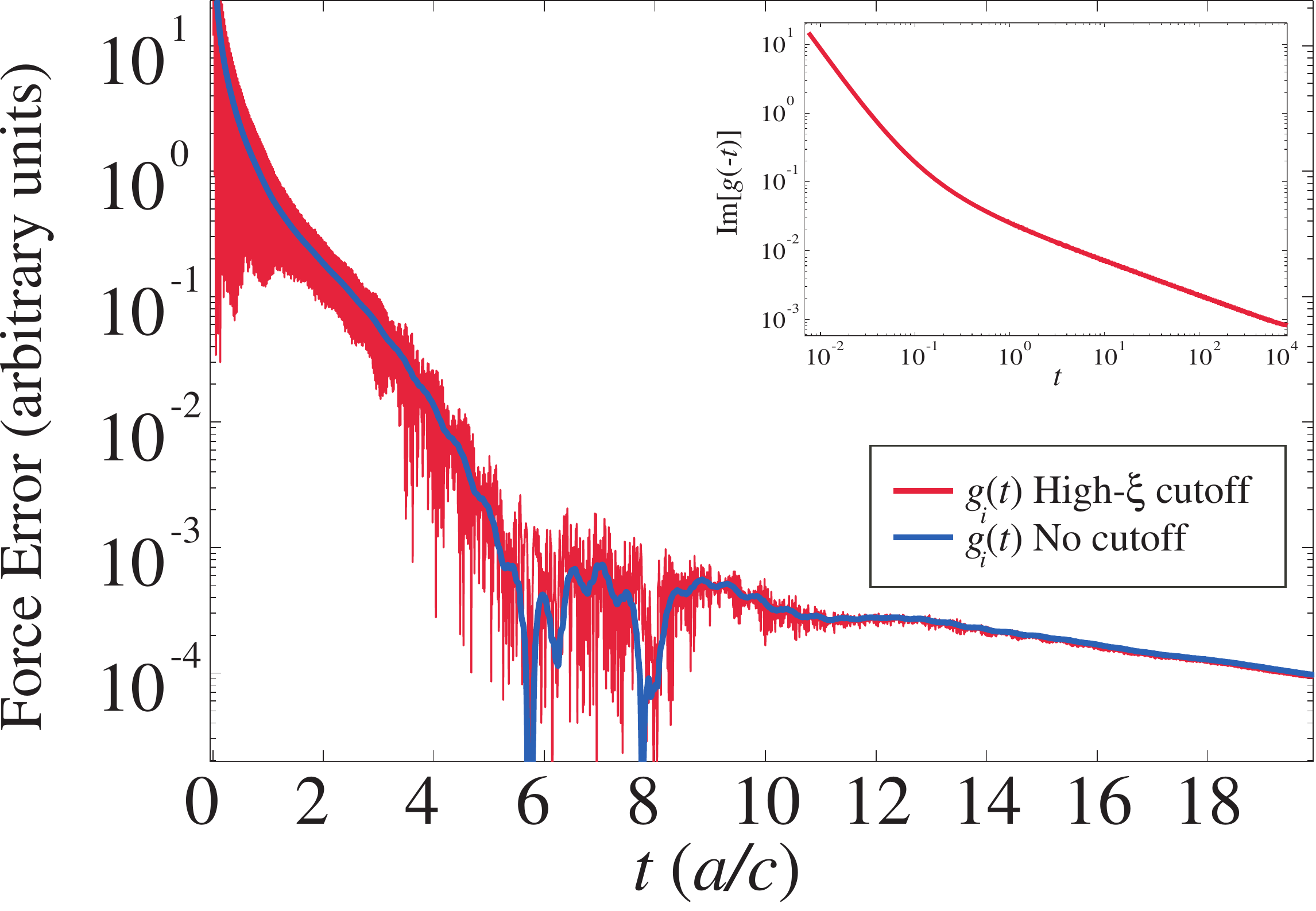}
\centering
\caption{Plot of the force error (force after a finite time
  integration vs. the force after a very long run time) for $g(t)$
  determined from a numerical transform as
  in~\citeasnoun{RodriguezMc09:PRA} and from the analytic transform of
  the high-frequency components.  Inset: $\Im[g(-t)]$ obtained without
  a cutoff, in which the high-frequency divergence is integrated
  analytically.  Compare with Fig. 1
  of~\citeasnoun{RodriguezMc09:PRA}}
\label{fig:gt-new}
\end{figure}

In~\figref{gt-new} we plot the convergence of the force as a function
of time for the same system using the $g(-t)$ obtained by use of a
high-frequency cutoff and for one in which $g_1(\xi)$ is transformed
analytically and the remainder is transformed without a cutoff.  The
inset plots $\Im g(-t)$ obtained without using a cutoff (since the
real part is not used in this paper) for $\sigma=10$.  If a complex
harmonic basis is used, one must take care to use the full $g(t)$ and
not only its imaginary part.

\subsubsection{Further simplification}

In addition to the treatment of the high-frequency divergence in the
previous section, we find it convenient to also Fourier transform the
low-frequency singularity of $g(\xi)$ analytically.  As discussed
in~\citeasnoun{RodriguezMc09:PRA}, the low-frequency limit of $g(\xi)$
is given by:
\begin{equation}
g(\xi) \rightarrow g_2(\xi)\equiv \frac{\sqrt{i}}{2} \frac{\sigma^{3/2}}{\xi^{1/2}} \Theta(\xi)
~\mathrm{as}~\xi\rightarrow 0
\end{equation}

The Fourier transform of $g_2(\xi)$, viewed as a distribution, is:
\begin{equation}
g_2(-t) = \frac{i}{4\sqrt{\pi}} \frac{\sigma^{3/2}}{t^{1/2}}
\end{equation}

After removing both the high- and low-frequency divergences of
$g(\xi)$, we perform a numerical Fourier transform on the function
$\delta g(\xi) \equiv g(\xi) - g_1(\xi) - g_2(\xi)$, which is
well-behaved in both the high- and low-frequency limits. 

In the present text we are only concerned with real sources, in which
case all fields $\Gamma(\vec{x},t)$ are real and only the imaginary
part of $g(-t)$ contributes to the force in~\Eqref{time-force}.  The
imaginary part of $g(-t)$ is then:

\begin{equation}
\Im[g(-t)] = \Im (\delta g(-t)) + \frac{1}{2\pi}\left(\frac{1}{t^2} + \frac{\sigma}{t}\right)
+ \frac{1}{4\sqrt{\pi}} \frac{\sigma^{3/2}}{t^{1/2}}
\end{equation}

\subsubsection{Perfect conductors and $z$-invariance}

As discussed in~\citeasnoun{Rodriguez07:PRA}, the stress-tensor
frequency integral for a three-dimensional $z$-invariant system
involving only vacuum and perfect metallic conductors is identical in
value to the integral of the stress tensor for the associated
two-dimensional system (corresponding to taking a $z=0$ crossection),
with an extra factor of $i\omega/2$ in the frequency integrand.  In
the time domain, this corresponds to solving the two-dimensional system
with a new $g(-t)$.

In this case the Fourier transform can be performed analytically.  The
result is:

\begin{equation}
\Im [g(-t)] = \frac{1}{2\pi} \left( \frac{2}{t^3} + \frac{3 \sigma}{2t^2} + \frac{\sigma^2}{2 t}\right)
\end{equation}

%Note that since $S$ is a closed surface, the integral of $\Gamma(x,t)$
%over $S$ for very short times is always zero, so the $t\rightarrow 0$
%divergence in $g(t)$ above will not play a role in the force.

\subsection{Harmonic expansion in cylindrical coordinates}
\label{sec:app-mult}

The extension of the above derivation to three dimensions and
non-Cartesian coordinate systems is straightforward, as the only
difference is in the representation of the $\delta$-function.  Because
the case of rotational invariance presents some simplification, we
will explicitly present the result for this case below.

For cylindrical symmetry, we work in cylindrical coordinates
$(r,\phi,z)$ and choose a surface $S$ that is also rotationally
invariant about the $z$-axis.  $S$ is then a surface of revolution,
consisting of the rotation of a parametrized curve
$(r(s),\phi=0,z(s))$ about the $z$ axis.  The most practical harmonic
expansion basis consists of functions of the form $f_n(x) e^{i m
  \phi}$.  Given a $\phi$ dependence, many FDTD solvers will solve a
modified set of Maxwell's Equations involving only the $(r,z)$
coordinates.  In this case, for each $m$ the problem is reduced
to a two-dimensional problem where both sources and fields are
specified only in the $(r,z)$-plane.

Once the fields are determined in the $(r,z)$-plane, the force
contribution for each $m$ is given by:
\begin{multline}
\int_0^{2\pi} d\phi\int_S ds_j(\vec{x})\, r(\vec{x}) e^{-im\phi} \\ 
\, \int_0^{2\pi} d\phi^\prime \int_S ds(\vec{x}^\prime)\,
r(\vec{x}^\prime) e^{im\phi^\prime}
\delta_S(\vec{x}-\vec{x}^\prime)\Gamma_{ij;m}^E(t;\vec{x},\vec{x}^\prime)
\label{eq:cyl-force-m}
\end{multline}
where the values of $\vec{x}$ range over the full three-dimensional
$(r,\phi,z)$ system.  Here we introduce the Cartesian line element
$ds$ along the one-dimensional surface $S$ in anticipation of the
cancellation of the Jacobian factor $r(\vec{x})$ from the integration
over $S$.  We have explicitly written only the contribution for $\Gamma^E$,
the contribution for $\Gamma^H$ being identical in form.

In cylindrical coordinates, the representation of the $\delta$-function
is:
\begin{equation}
\delta(\vec{x}-\vec{x}^\prime) 
= \frac{1}{2\pi r(\vec{x})} \delta(\phi-\phi^\prime)\delta(r-r^\prime)\delta(z-z^\prime)
\end{equation}

For simplicity, assume that $S$ consists entirely of $z=$ constant and
$r =$ constant surfaces (the more general case follows by an analogous
derivation).  In these cases, the surface $\delta$-function $\delta_S$
is given by:
\begin{eqnarray*}
\delta_S(\vec{x}-\vec{x}^\prime) = \frac{1}{2\pi r(\vec{x})} \delta(\phi-\phi^\prime)\delta(r-r^\prime),~~~z=\mathrm{constant} \\
\delta_S(\vec{x}-\vec{x}^\prime) = \frac{1}{2\pi r(\vec{x})} \delta(\phi-\phi^\prime)\delta(z-z^\prime),~~~r=\mathrm{constant}
\end{eqnarray*}
In either case, we see that upon substitution of either form of
$\delta_S$ into \eqref{cyl-force-m}, we obtain a cancellation with the
first $r(\vec{x})$ factor. Now, one picks an appropriate decomposition of
$\delta_S$ into functions $f_n$ (a choice of $r=$ const or $z =$ const
merely implies that the $f_n$ will either be functions of $z$, or $r$,
respectively). We denote either case as $f_n(\vec{x})$, with the $r$
and $z$ dependence implicit.

We now consider the contribution for each value of $n$.  The integral
over $\vec{x}^\prime$ is:
\begin{multline*}
\Gamma^E_{ij;nm}(t,\vec{x}) = \\
\int_0^{2\pi} d\phi^\prime \int_S ds(\vec{x}^\prime) r(\vec{x}^\prime)
\Gamma^E_{ij;nm}(t,\vec{x},\vec{x}^\prime) f_n(\vec{x}^\prime) e^{im\phi^\prime}
\end{multline*}

As noted in the text, $\Gamma^E_{ij;nm}(t,\vec{x})$ is simply the
field measured in the FDTD simulation due to a three-dimensional
current source of the form $f_n(\vec{x})e^{im\phi}$.  In the case of
cylindrical symmetry, this field must have a $\phi$ dependence of the
form $e^{im\phi}$:
\begin{equation}
\Gamma^E_{ij;nm}(t,r,z,\phi) = \Gamma^E_{ij;nm}(t,r,z)e^{im\phi}
\end{equation}

This factor of $e^{im\phi}$ cancels with the remaining $e^{-im\phi}$.
The integral over $\phi$ then produces a factor of $2\pi$ that cancels
the one introduced by $\delta_S$.  After removing these factors, the
problem is reduced to one of integrating the field responses entirely
in the $(r,z)$ plane.  The contribution for each $n$ and $m$ is then:
\begin{equation}
\int_S ds_j(\vec{x})\, \bar{f}_n(\vec{x}) \Gamma^E_{ij;nm}(t,r,z)
\end{equation}

If one chooses the $f_n(\vec{x})$ to be real-valued, the contributions
for $+m$ and $-m$ are related by complex conjugation.  The sum over
$m$ can then be rewritten as the real part of a sum over only non
negative values of $m$.  The final result for the force from the
electric field terms is then:
\begin{equation}
F_i = \int_0^\infty dt \,\Im[g(-t)]\sum_{n} \int_S ds_j(r,z)\, f_n(r,z) \Gamma^E_{ij;n}(t,r,z)
\end{equation}
where the $m$-dependence has been absorbed into the definition of
$\Gamma_{ij;n}$ as follows:
\begin{equation}
\Gamma^E_{ij;n}(t,r,z) \equiv \Gamma^E_{ij;n,m=0}(t,r,z)
+ 2\sum_{m>0}\Re[\Gamma^E_{ij;nm}(t,r,z)]
\end{equation}

We have also explicitly included the dependence on
$r$ and $z$ to emphasize that the integrals are confined to the
two-dimensional $(r,z)$ plane.  The force receives an analogous
contribution from the magnetic-field terms.

%\bibliographystyle{apsrev} 
%\bibliography{photon}

\end{document}